\RequirePackage{xcolor} 

\documentclass[journal,twoside,web]{ieeecolor}

\usepackage{etoolbox}
\makeatletter
\@ifundefined{color@begingroup}%
  {\let\color@begingroup\relax
   \let\color@endgroup\relax}{}%
\def\fix@ieeecolor@hbox#1{%
  \hbox{\color@begingroup#1\color@endgroup}}
\patchcmd\@makecaption{\hbox}{\fix@ieeecolor@hbox}{}{\FAILED}
\patchcmd\@makecaption{\hbox}{\fix@ieeecolor@hbox}{}{\FAILED}

\usepackage{jsen}
\usepackage{cite}
\usepackage{amsmath,amssymb,amsfonts}
\usepackage{algorithmic}
\usepackage{graphicx}
\usepackage{textcomp}
\usepackage{wrapfig}
\usepackage{textcomp}
\usepackage{comment}
\usepackage[T1]{fontenc}
\usepackage{float}

\usepackage{tikz}
\usepackage[hidelinks]{hyperref}
\definecolor{lime}{HTML}{A6CE39}
\DeclareRobustCommand{\orcidicon}{
	\begin{tikzpicture}
	\draw[lime, fill=lime] (0,0) 
	circle [radius=0.16] 
	node[white] {{\fontfamily{qag}\selectfont \tiny ID}};
	\draw[white, fill=white] (-0.0625,0.095) 
	circle [radius=0.007];
	\end{tikzpicture}
	\hspace{-2mm}
}
\foreach \x in {A, ..., Z}{\expandafter\xdef\csname orcid\x\endcsname{\noexpand\href{https://orcid.org/\csname orcidauthor\x\endcsname}
			{\noexpand\orcidicon}}
}
\def\BibTeX{{\rm B\kern-.05em{\sc i\kern-.025em b}\kern-.08em
    T\kern-.1667em\lower.7ex\hbox{E}\kern-.125emX}}
\definecolor{abstractbg}{rgb}{0.89804,0.94510,0.83137}
\begin{document}

    \title{Si Superstrate Lenses on Patch-Antenna-Coupled TeraFETs: NEP Optimization and Frequency Fine-Tuning}

\author{Anastasiya Krysl, 
Dmytro B. But\orcidB{}, 
Kęstutis~Ikamas\orcidD{}, 
Jakob~Holstein\orcidI{}, 
Anna Shevchik-Shekera\orcidE{}, 
Hartmut G. Roskos\orcidF{}
and
Alvydas Lisauskas \orcidC{}, \IEEEmembership{Member, IEEE} 
\thanks{A. Krysl, J. Holstein, and H. G. Roskos are with the Physi\-ka\-li\-sches Institut, Goethe University, 60438 Frankfurt am Main, Germany (email addresses: krysl@physik.uni-frankfurt.de; holstein@physik.uni-frankfurt.de; roskos@physik.uni-frankfurt.de).}
\thanks{D. B. But is with CENTERA Laboratories, Institute of High Pressure Physics PAS, Warsaw, 01-142 Poland, and with NOMATEN Centre of Excellence, National Centre of Nuclear Research, 05-400 Otwock-Świerk, Poland (e-mail: dbut@unipress.waw.pl).}
\thanks{K. Ikamas is with Institute of Applied Electrodynamics and Telecommunications, Vilnius University, LT-10257 Vilnius, 10257, Lithuania, and also with General Jonas \v{Z}emaitis Military Academy of Lithuania, LT-10322 Vilnius, Lithuania (email: kestutis.ikamas@ff.vu.lt).}
\thanks{A. Shevchik-Shekera is with V.Ye. Lashkaryov Institute of Semiconductor Physics (ISP), NASU, Kyiv, 03028, Ukraine (email: shevchik-shekera@isp.com.ua).}
\thanks{A. Lisauskas is with Physikalisches Institut, Goethe University Frankfurt, 60438 Frankfurt am Main, Germany, and with the Institute of Applied Electrodynamics and Telecommunications, Vilnius University, LT-10257 Vilnius, Lithuania, (email: alvydas.lisauskas@ff.vu.lt).}
}

\IEEEtitleabstractindextext{%
\fcolorbox{abstractbg}{abstractbg}{%
\begin{minipage}{\textwidth}%
\begin{wrapfigure}[12]{r}{3in}%
\includegraphics[width=2.15in, height=1.65in]{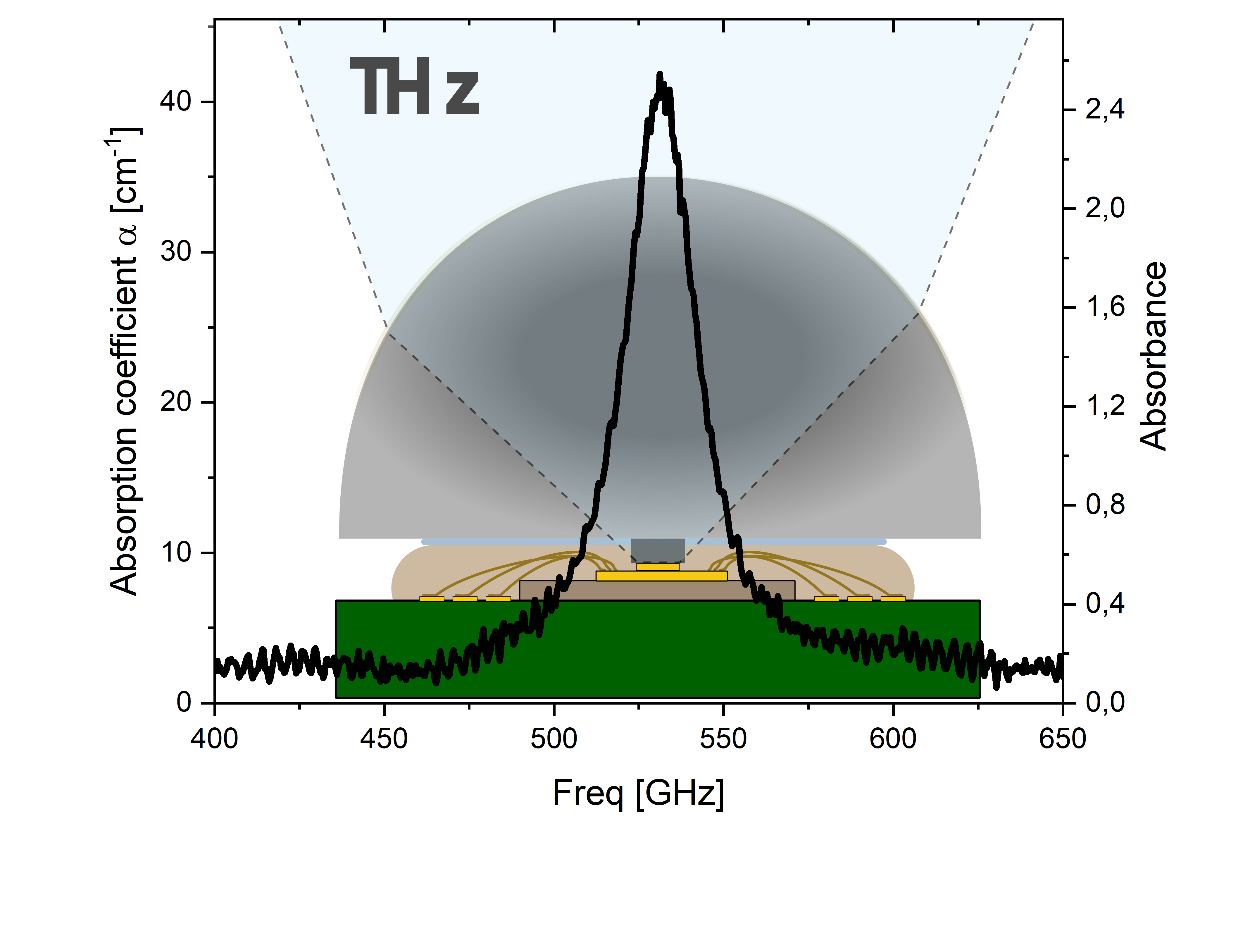}
\end{wrapfigure}%
\begin{abstract}
This paper presents a study on performance optimization and resonant frequency modification of terahertz detectors by the use of hyper-hemispherical silicon superstrate lenses. The detectors are patch-TeraFETs, i.e., field-effect transistors with monolithically integrated patch antennas fabricated with a commercial 65-nm CMOS foundry process and designed for an operation frequency of 580~GHz. We demonstrate a strong improvement of the optical noise-equivalent power (\textit{optical NEP}, referenced against the total radiation power) reaching a value of 16~pW/$\sqrt{Hz}$. We show furthermore, that the resonance frequency 
can be efficiently fine-tuned by the choice of the material and the thickness of a dielectric layer placed between the transistor and the superstrate lens. The resonance frequency can be shifted by more than 15\% of the center frequency (up to 100~GHz for the 580~GHz devices). The design of the on-chip optics can be employed for post-fabrication tailoring of the detector's resonance frequency to target specific spectral positions. 
\end{abstract}

\begin{IEEEkeywords}
CMOS, field-effect transistor, terahertz radiation, THz detection, superstrate lens  
\end{IEEEkeywords}
\end{minipage}}}

\maketitle

\section{Introduction}
\label{sec:introduction}
\IEEEPARstart
The accelerating growth of THz-based technologies is unlocking a multitude of new applications in areas such as security monitoring, wireless communications, medical imaging, quality control and more\cite{Valusis2021, Singh2020, Alqaraghuli2023, Yan2022, Ahi2016}. 
The potential for many of these applications arises by specific material properties, either because materials are (semi-)transparent in the THz frequency range, or because they interact in a pronounced way with the radiation and exhibit  unique spectral properties in the THz range. These interactions occur with the material's internal degrees of freedom, including vibrational and rotational modes as well as electronic excitations, which makes THz radiation valuable for the identification of substances' spectral fingerprints in various contexts \cite{Naftaly2007, Bauer2014, Fu2022, Rothbart2022, Warawa2023}.
Moreover, THz radiation is non-ionizing because of the low photon energy (a frequency of 1~THz corresponding to a photon energy of about 4.1~meV), which avoids health risks in applications. 

An interesting example of a potential application is the detection of water molecules in air or in other gaseous environments, since it can be identified by strong rotational transition lines, one of which is located at 557~GHz~\cite{HITRAN2020}. This characteristic absorption line could be utilized for studies of water in the atmosphere or 
for air humidity control in the food industry or in intensive-care stations in hospitals, 
where noninvasive measurements methods are in high demand. 
Another substance of interest, which has an absorption signature in this frequency range, is alpha-lactose monohydrate. It is obtained as the crystalline form of lactose from the aqueous solution. This lactose anomer can be detected by its vibrational mode at 530~GHz \cite{Brown_2007}. Its rapid identification can be useful 
in the dairy and infant-nutrition industries and for  medical diagnosis\cite{Churakova2019, DaSilva_2022}. 
In the pharmaceutical industry, where lactose often serves as binder in pills \cite{Hebbink2019}, it is important to ensure that isomerically pure lactose is used, as the $\alpha$- and $\beta$-anomers 
vastly vary 
in solubility, compressibility, and reactivity, which can impact drug stability and manufacturing processes.  



In the last decade, one could notice a growing increase in the use of field-effect-transistors (FETs) as devices for the detection of THz radiation. For this purpose, they are commonly embedded monolithically in antenna structures and are then called TeraFETs. The idea of using FETs for the detection of THz radiation was first introduced in \cite{Dyakonov1993}, where it was suggested to employ the nonlinear dependence of the drain-source current on the gate bias voltage for the rectification of an AC current induced by the incoming radiation. 
The concept became technologically relevant following the carefully designed incorporation of antennas \cite{Lis09, Ojefors2009}.  TeraFETs have several advantages over other types of THz detectors, including high sensitivity, fast response time, low noise and compactness\cite{Marczewski2018}. They cover a large part of the THz spectral regime \cite{Boppel2012} and can be used as power detectors or heterodyne receivers \cite{LisauskasSubharmonic}. They are fabricated using standard (foundry) semiconductor manufacturing methods, thus being produced with high yield and low performance variations \cite{Boppel2011}, and becoming relatively inexpensive compared to other types of THz detectors which require specialized fabrication techniques. Last but not least, the foundry-based processing allows research institutions to benefit from advanced fabrication technologies at affordable costs and within reasonable production time frames.

The choice of the type of antenna to be integrated and its design depend on the targeted frequency ranges, application requirements, and restrictions imposed by the fabrication process. 
With regard to the latter, it is important to note that one has to strictly adhere to the design rules and usually tight specifications of the foundry. In order to pass the (often changing) design rules check (DRC), a detector designer 
may encounter significant difficulties on the way to achieve state-of-the-art integrated-antenna performance, such that the task may require considerable intuition and its handling becomes a form of art. 

A type of antenna, which alleviates some of the challenges of integration specifically for Si CMOS-type detectors, is the patch antenna \cite{Lis09,Ojefors2009, Boppel2011}. The back end of line of the IC process technology, which offers the possibility to deposit and pattern more than ten layers of metal and oxide dielectric for the fabrication of interconnects and passive circuit components (capacitors, inductors), represents a natural platform for the implementation of the antenna patch and the ground-plane -- both separated by up to about 12~$\mu$m of dielectric -- and of the via interconnects needed to bring the THz signal from the patch to the FET.  

We note in passing that at lower frequencies, in the microwave spectral region, the microstrip patch antenna stands out as one of the most prevalent antenna designs. Its appeal arises from such qualities as its simple topological profile, low weight, ease of fabrication, suitability for array implementation, and ease of connection with microstrip transmission lines. It exhibits a fairly small bandwidth at its center frequency. Microwave applications that make use of microstrip patch antennas encompass automotive radar for object detection and collision avoidance \cite{Patole2017}, wireless backhaul links for high-capacity data transmission \cite{Sonkki2018}, 6G wireless communication network \cite{Nissanov2023}, and breast cancer detection \cite{Geetharamani2019}. 

The patch antenna has, however, a major drawback. The fact, that the electric field pattern of the antenna extends from the patch to the buried ground-plane through the back end's dielectric with its dielectric constant $\varepsilon \gg 1$, leads to a physical size of the patch (in the direction of the polarization of the radiation) which is smaller than half a wavelength of the radiation in free space. The effective antenna cross-section (which typically is slightly larger than the physical size of the patch \cite{Ojefors2009}) is hence not well matched to the spot size of the THz radiation incoming from free space even if the radiation is focused by a lens or a mirror. This results in a poor coupling efficiency, as a substantial amount of the power of the beam bypasses the antenna. 

With ground-plane-free types of antennas, where a similar challenge exists because the local field extends into the substrate, one uses a substrate lens attached to the backside of the detector chip, to focus the radiation tightly to match the effective antenna cross-section, but this is not possible in the case of the patch antenna because the buried ground-plane prevents the use of a substrate lens. On the other hand, attachment of a superstrate lens to the detector's front-side is hindered by the non-planar topography of the detector's surface, but much more so by the bonding wires used for electric contacting of the device. 

Here, we introduce the use of a Si superstrate lens in combination with a composite dielectric buffer layer for surface smoothing and the embedding of the bonding wires. With this development, we extend earlier work, where superstrate lenses entirely made from paraffin wax were studied and employed in THz imaging systems \cite{Yuan19,yuan2023a,yuan2023b}. The new radiation-focussing technology presented now is characterized by a high reproducibility and reliability as well as a low insertion loss of the optical components. We show that we achieve efficient front-side radiation coupling. We also demonstrate that the choice of the composition of the buffer layer 
allows to dielectrically fine-tune the frequency of peak sensitivity 
of the TeraFETs. 

We showcase these developments with 
a patch-antenna-coupled detector designed for operation at 580~GHz. The device, when equipped with a suitable superstrate assembly, is suitable at the same time for the measurement of the absorption lines of lactose at 530~GHz and of water vapor in air at 557~GHz.


\section{Design and fabrication}
\subsection{Die Fabrication}
\begin{figure}[!h]
  \centering
  \includegraphics[width=0.95\linewidth]{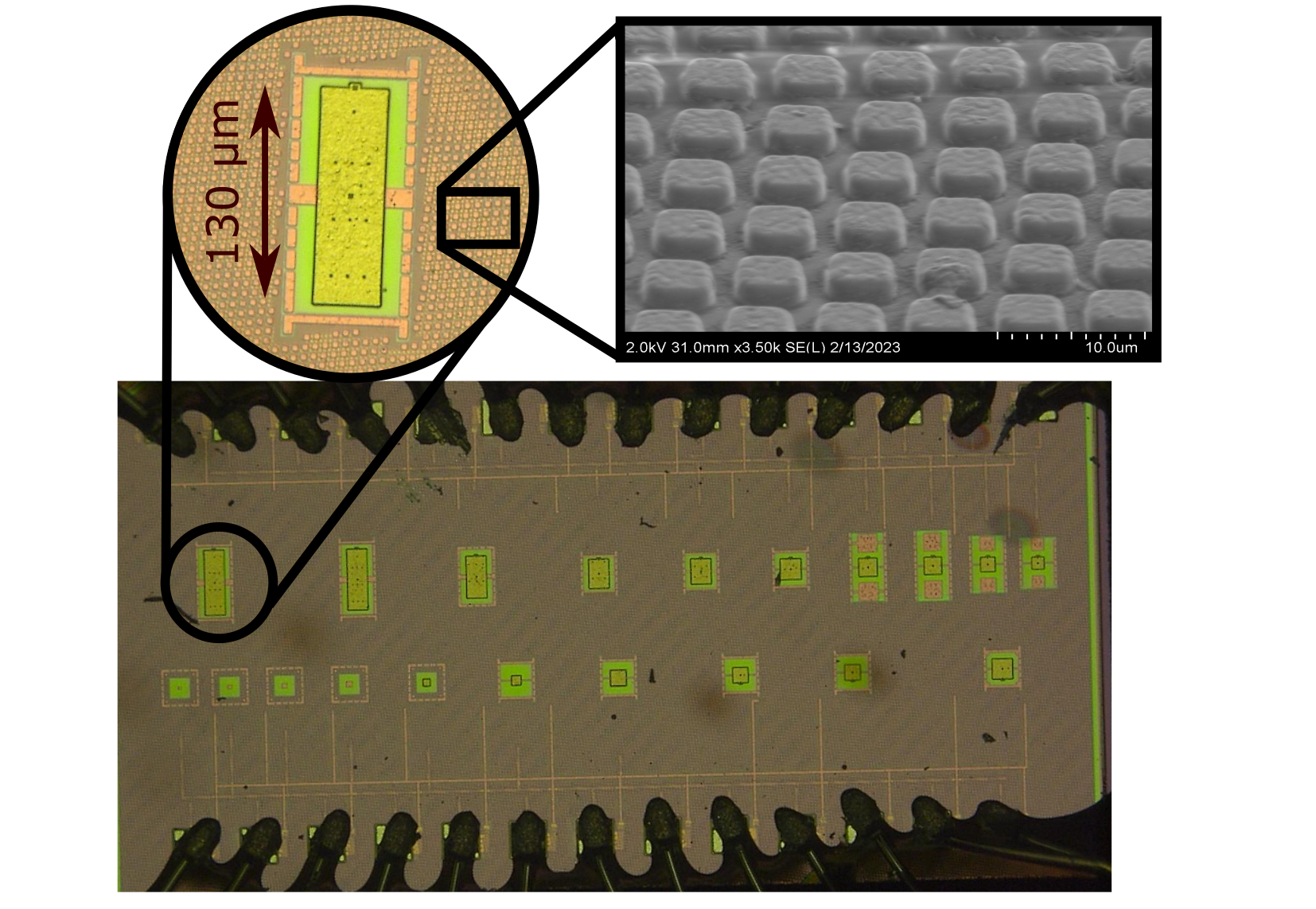}
  \caption{Detector die as delivered by the foundry. It contains two rows of patch-antenna-coupled TeraFETs, with ten devices in each row. The patch antennas of the TeraFETs have different sizes and shapes. Image on top left: A detailed view of the detector investigated in this paper. Image on top right: SEM image on the die's surface showing the regular array of metal pillars incorporated by the foundry for strain relief.}
\label{fig:DetectorDieOverview}
\end{figure}

For this work, TeraFETs were fabricated at the TSMC foundry in Taiwan using their  65-nm CMOS technology node. The Si substrate thickness was 270 $\mu$m. Fig.~\ref{fig:DetectorDieOverview} shows a photograph of an exemplary die carrying 
patch-antenna-coupled TeraFETs with different sizes of the patches and hence also different resonance frequencies in the range from 0.5 to 2.5~THz.
In addition, a magnified view on the detector investigated in this article is shown, as well as a scanning-electron-microscope (SEM) image of the die's surface. The SEM picture displays the regular surface pattern of the metallic islands required for strain relief by the foundry's design rules.
The investigated detector contained a single FET with a channel length and width of 60~nm and 1~$\mu$m, respectively. The antenna patch was implemented in the top metal layer (out of 10 total). Its thickness was 1.45~$\mu$m and the separation $h$ between the patch and the ground-plane is 8.41~$\mu$m. The width $W$ and length $L$ of the patch were 40~$\mu$m and 130~$\mu$m, respectively. 
The THz radiation was polarized along the length of the patch.

\subsection{Measurement Cases}

\begin{figure}[t]
  \centering
 \includegraphics[width=0.95\linewidth]{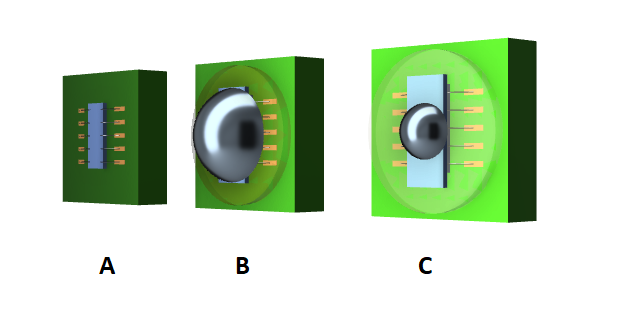}
 (a)\includegraphics[width=0.95\linewidth]{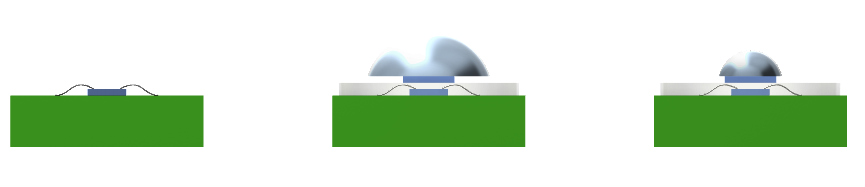}
 (b)\includegraphics[width=0.95\linewidth]{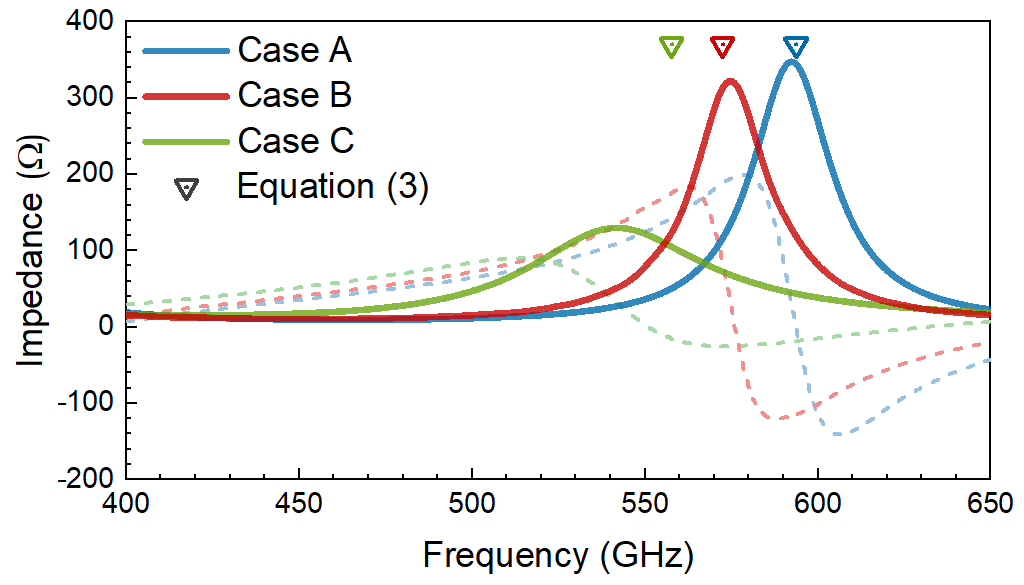}
  \caption{
  a) Schematic representation of the detector modules in the three different implementations A to C discussed in the text (the detector elements are not drawn to scale). b) Simulated real part (solid lines) and imaginary part (dashed lines) of the impedance of the planar antenna for the three implementations A to C. The triangular symbols mark the antenna resonance frequencies calculated with Eqs.~(\ref{eq:effective_eps})-(\ref{eq:res_freq}). 
  \\ \\ 
  }
 \label{fig:Cases&Impedance}
\end{figure}

Two devices of the same kind of antenna-coupled Tera\-FET as described above were equipped with superstrate lenses using two different approaches and materials. A third one remained umodified and served as reference. All three detectors are schematically shown in Fig.~\ref{fig:Cases&Impedance}. 
They will be addressed from now on as Case A, Case B, and Case C.

All three devices were mounted on PCB boards and wire-bonded for electrical contacting. Case A is the reference device which was not processed further. 
Case B is a detector whose contacts and wire bonds were embedded in an EPO-TEK\textsuperscript{\textregistered} 353ND epoxy to avoid damage to them upon further processing. The epoxy used is a EPO-TEK 353ND is a two-component, thermally curable high temperature epoxy designed for semiconductor applications. As shown in Fig.~\ref{fig:DetectorDieOverview}, the contact pads with the wire bonds had a fairly large separation from the two rows of detectors. This allowed us to apply the epoxy only to the region  with the wires, but to leave the area above the patch antennas uncovered. The antennas were hence located in a trench formed by the epoxy (see schematic illustration in the cross-sectional plots of Fig.~\ref{fig:Cases&Impedance}(a)).

The trench was subsequently filled with paraffin wax by drop-casting of molten wax (for handling details, see \cite{yuan2023b}). This produced an approximately 1-mm-thick stripe of wax above the rows of antennas. After the wax had cooled down, high-viscosity silicone paste (vacuum grease) was applied on top of the epoxy and the wax, and a Si lens (12-mm diameter and 6.8-mm height) was softly pressed onto the stack. The silicone paste served as a soft filler between the wax-covered detectors and the Si lens. 
In Case C, we left away the paraffin wax (which has non-negligible absorption losses at THz frequencies \cite{yuan2023b}). Instead, we applied a small amount of silicone paste in the trench left open in the epoxy, and placed a 0.4-mm-thick planar slab of Si into the trench. On top of it, we applied more silicone paste, onto which a Si lens (diameter: 4~mm, height: 2.2~mm) was attached. In both Case B and Case C, the viscous silicone grease holds the lens in place but still allows for smooth sliding of the lens upon alignment of the lens relative to the TeraFET. This alignment was performed while illuminating the detector with THz radiation and maximizing the rectified voltage by fine-tuning the position of the lens.


\subsection{ Antenna Design and Analysis}

\begin{figure}[!h]
  \centering
  (a)\includegraphics[width=0.50\linewidth]{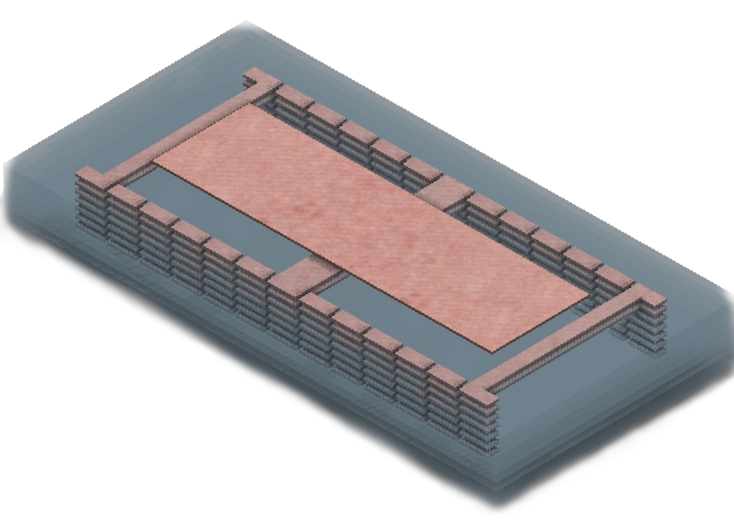}
(b)\includegraphics[width=0.95\linewidth]{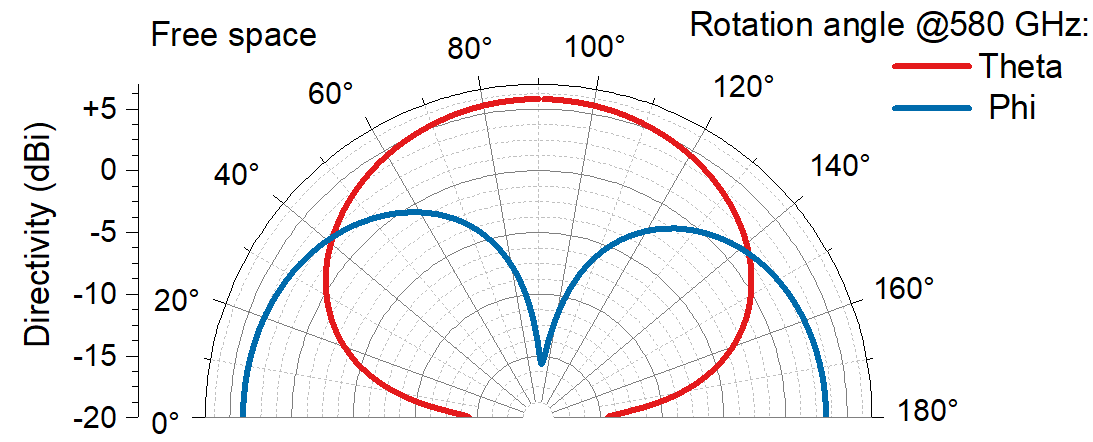}
  \caption{
  a) Schematic drawing of the investigated patch antenna with all additional inserted metal blocks required to fulfill density requirements. The via to the FET and the ground plane are not shown.
  b) Radiation pattern of the integrated patch antenna at 570~GHz which results in a calculated directivity of 5.8~dBi at the maximum. 
  \\ 
  }
 \label{fig:structure}
\end{figure}

An extensive analysis was carried out to examine the antenna characteristics thoroughly in dependence of the structure and the dielectric environment of the antenna. 
With regard to the latter, the different dielectric materials within the antenna and above it must be accounted for in the antenna simulations. This requires either an exact treatment of all layers or the implementation of estimation methods as described in \cite{Schneider1969, But2020}. We followed a two-step approach. First, we  considered the multi-layer stack of dielectrics within the antenna, i.e., between the ground-plane and the patch. 
We approximated the stack as a homogeneous isotropic layer with the equivalent permittivity of~\cite{Ali98}
\begin{equation}
\varepsilon_{eq} = \left (\sum^{N}_{n=1} \frac{h_n}{\varepsilon_{r n}}\right )^{-1}  \cdot \sum^{N}_{n=1}h_n,
\label{eq:effective_eps}
\end{equation}
where $h_n$ and $\varepsilon_{r n}$ are the height and the permittivity of the $n$th dielectric layer. 
The TSMC 65nm process utilizes 9 dielectric layers, and detailed information on layer thickness and dielectric properties is accessible through an NDA with the foundry. Due to confidentiality, specific layer details are not disclosed in this paper.
In our case, the equivalent permittivity $\varepsilon_{eq}$ for the used fabrication technology amounts to 4.04. 

In a second step, we considered the fringing of the electric field to the region above the patch, where the dielectric materials applied for the attachment of the substrate lens influence the effective permittivity. The latter can be estimated as proposed in \cite{Schneider1969}. The effects of both the fringing of the field 
and the existence of the superstrate material can be treated as an increased effective width $W_{eff}=W+2\Delta W$ and an effective length $L_{eff}=L+2\Delta L$ of the patch, with 

\begin{equation}
    \Delta W = \frac{h}{\pi}\ln\left(17.08\left(\frac{W}{2h} + 0.92\right)\right) \,,
\end{equation}
\begin{equation}
    \Delta L = 0.412\cdot h\frac{\varepsilon_{eff}+0.3}{\varepsilon_{eff}-0.259} \,,
\end{equation}
and the effective dielectric constant
\begin{equation}
\varepsilon_{eff}=\frac{\varepsilon_{eq}+\varepsilon_{s}}{2}+\frac{\varepsilon_{eq}-\varepsilon_{s}}{2}\left(\sqrt{1+\frac{12 h}{W}}\right)^{-1} \,,
\label{eq:effective_eps_res}
\end{equation}
where $h$ is the distance between the ground-plane and the patch, $W$ is the width of the patch, and $\varepsilon_{s}$ is the dielectric permittivity of the composite material used as superstrate. 

The resonance frequency of the antenna was estimated with the following formula:
\begin{equation}
{f}=\frac{c}{2(L+\Delta L) \sqrt{\varepsilon_{eff}}} \,.
\label{eq:res_freq}
\end{equation}
For simplicity, we assumed that $\Delta L=h$ which equals the distance from the patch to the ground plane  

Eq.~(\ref{eq:res_freq}) allowed us to estimate the resonance frequency for the case without the superstrate (Case A), obtaining a value of 594~GHz. For Case B, we assumed that the effective dielectric constant of the superstrate could be approximated by that of the paraffin wax ($\varepsilon_s=2.1$), neglecting a possible influence of the silicone grease and the Si substrate lens because of the strong confinement of the antennas near-field to the patch. The resultant resonant frequency is  573~GHz. For case C, we similarly assumed the dielectric constant to be approximatively that of the used epoxy ($\varepsilon_s=2.95$), which led to a predicted resonance at 557~GHz. 

The bare antenna structure (Case A) with the metal components as shown in  Fig.~\ref{fig:structure}(a)) was analyzed using the CST Studio Suite software (Dassault Systèmes) using a time domain solver. 
The simulations were performed over the frequency range from 300~GHz to 700~GHz. The blue solid and dashed curves in Fig.~\ref{fig:Cases&Impedance}(b) show the real and imaginary parts of the antenna impedance, respectively. 
At the resonance (590~GHz), the real part has a value of 340~$\Omega$. 
The simulated radiation efficiency is quite low and equals around 18~\%, and the directivity of the antenna is in the range of 5~dBi to 6.5~dBi within the simulated frequency range, amounting to 5.8~dBi at 580~GHz. Figure~\ref{fig:structure}(b) presents a conical projections of the radiation pattern, 
the angle theta, which is relative to the z-axis, and phi, which is relative to the x-axis in case the z-axis is perpendicular to the antenna face and x- is parallel to radiation polarization. 
\\

\subsection{Dielectric Loading Effect by the Superstrate Lens Assembly}
\begin{figure}[!h]
  \centering
  (a)\includegraphics[width=0.75\linewidth]{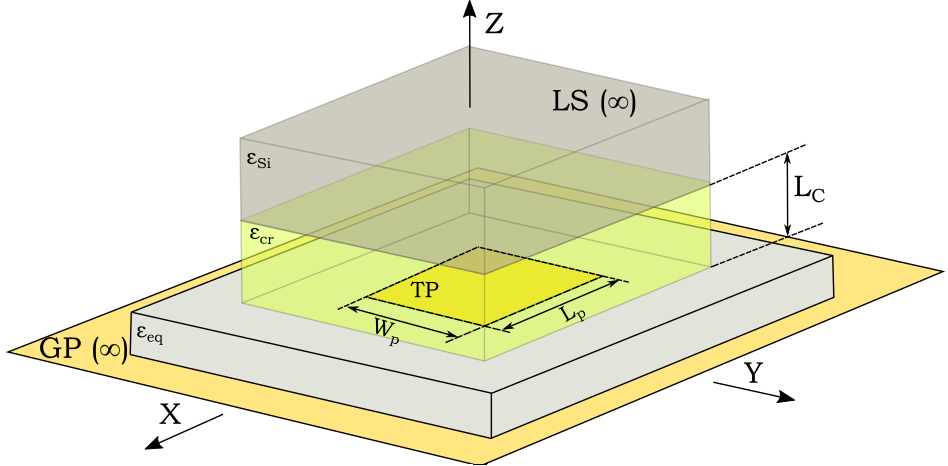}
  (b)\includegraphics[width=0.95\linewidth]{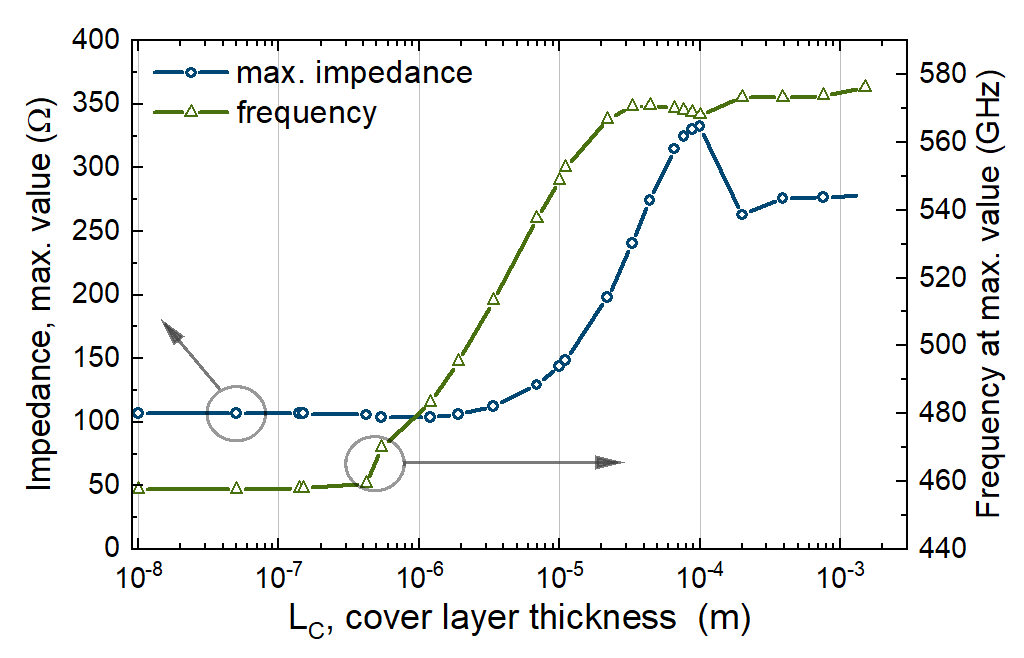}
  (c)\includegraphics[width=0.95\linewidth]{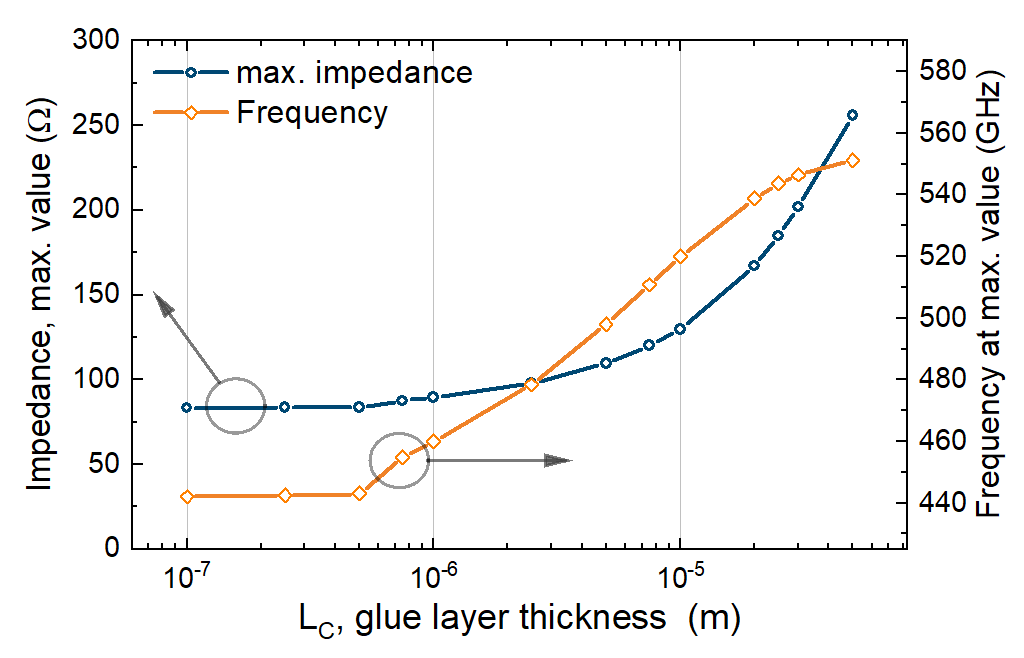}
  \caption{a) A model of the patch antenna with dielectric cover layers. 
  b) Resonance frequency and impedance (real part) as a function of the thickness of the paraffin cover layer (approximating Case B).
  c) Resonance frequency and impedance (real part) as a function of the thickness of the silicone glue layer (describing Case C).
  }
 \label{fig:CST}
\end{figure}

%
%
It is well established 
that employing a silicon hyper hemispherical lens attached to the die is an effective way to enhance the coupling efficiency of incoming free-space radiation to TeraFETs \cite{Bop16, Ikamas2021_homodyne}, as it conversely can provide an effective out-coupling of 
radiation from Si CMOS THz emitters and other emitters into free space 
\cite{Rudd02, Ikamas2021}. 
Before considering the radiation coupling in the next Section, 
we simulated the impedance and the resonance frequency of the antenna using the CST Studio Suite. For this purpose, we adopted a simplified antenna model with a cover layer composed of two materials as shown in Fig.~\ref{fig:CST}(a).
The insulator between the patch and the ground plane is characterized by the dielectric constant $\varepsilon_{eq}$, the equivalent permittivity of Eq.~(\ref{eq:effective_eps}). 
On top of the die, there is first a cover layer which represents the paraffin wax (Case B), respectively the silicone paste (Case C). The thickness is L$_C$ and the permittivity $\varepsilon_{cr}$. On top of the cover layer follows a layer of Si (marked as LS($\infty$) in the plot, with permittivity $\varepsilon_{Si}$), which represents the Si lens in Case B, respectively the Si slab plus the Si lens in Case C. 
The antenna itself was modeled as consisting of the top patch plate (TP) of width $\text{W}_P$ and length $\text{L}_P$, and of the extended ground-plane GP($\infty$). 

For the cover layer, we used the following material parameters: In case of paraffin wax, the literature specifies a permittivity in the range of 2.248–2.283, with a loss tangent ranging from 0.3~$\cdot10^{-3}$ to 7.7~$\cdot 10^{-3}$ ~\cite{yuan2023b, ghassemiparvin2016paraffin}. 
For Si, $\varepsilon_{Si} = 11.7$ and loss-less behavior. We note in passing that, in practical use, the power absorption coefficient of Si should be maintained below  0.05 cm$^{-1}$ to achieve the optimum performance for frequencies below 2~THz~\cite{Grischkowsky2004}.
The results of simulations by varying the thickness and material of the cover layer are presented in Fig.~\ref{fig:CST} (b) and (c). When the thickness of the paraffine layer exceeds 30~$\mu$m, the resulting resonant frequency of the patch reaches 570~GHz with slight oscillations of the maximum value of impedance due to the formation of a standing wave. For very thin layers, when the top silicon layer becomes in proximity to the antenna, the resonance frequency decreases to 460~GHz as well as the impedance at the resonance decreases to nearly 100~$\Omega$. Using a higher-loss material as a cover layer one can observe similar behavior with slightly lowered impedance and resonant frequency values as it is presented in Fig.~\ref{fig:CST} (c).

\section{Results of the THz Measurements and Discussion}

\begin{figure}[!h]
  \centering
  (a)\includegraphics[width=0.95\linewidth]{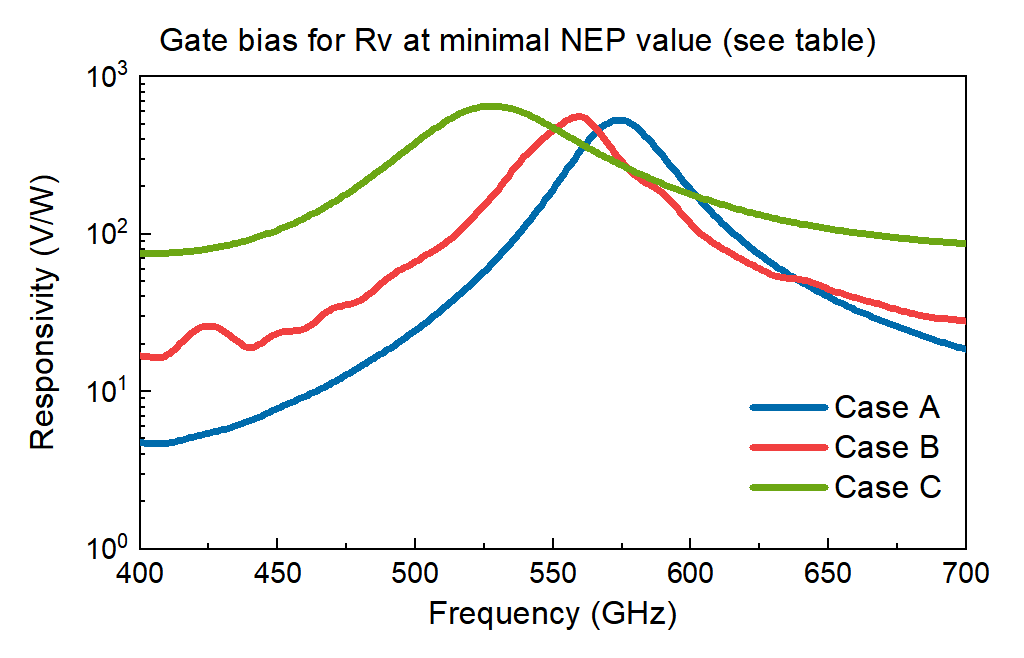}
  (b)\includegraphics[width=0.95\linewidth]{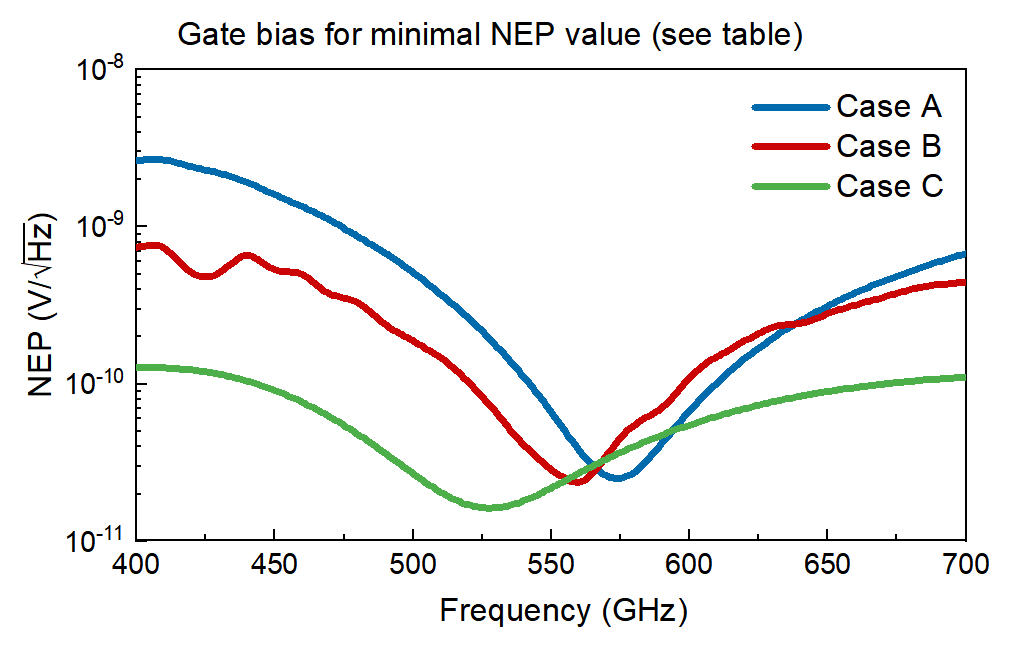}
  \caption{
  Frequency dependence of a) responsivity and b) NEP.
  }
\label{RespNep}
\end{figure}

The main experimental results are shown in Figs.~\ref{RespNep} and \ref{fig:image1} for a representative detector ``A1'' out of the full dataset taken of ten devices working at various frequencies. This device had a design resonance frequency of 580~GHz. The measured cross-sectional responsivity 
of the reference detector (Case A) is shown in Fig.~\ref{RespNep} as a blue line and indeed exhibits a peak in response at that design frequency. With that responsivity value, we calculated a cross-sectional NEP of 25~pW/$\sqrt{\text{Hz}}$ at 580~GHz. We point out here again, that care has to be taken when comparing this NEP value with those of the devices presented by Cases B and C, to be discussed in the next paragraph. Their NEP values are \textit{optical} NEP values. For the Case-A device, an optical NEP cannot reasonably be determined as much of the incident THz beam bypassed the detector as it had a comparatively low directivity of 6.6~dBi. 

We now consider the Case-B and Case-C detector assemblies. Both the cover layer(s) and the silicon lens added a dielectric load on the antenna which shifted its resonance frequency, but also improved both the radiation efficiency and the directivity. 

For Case B -- with the paraffin wax in the cover layer --, one finds in Fig.~\ref{RespNep} (red curve) a maximal optical responsivity of 559~V/W. The resonance frequency is down-shifted to 560~GHz. For the optical NEP, one calculates a value of 24~pW/$\sqrt{\text{Hz}}$. 

Concerning the sample of Case C, the improvement in the optical responsivity (green curve in Fig.~\ref{RespNep}) was even larger. A value of 656~V/W at the shifted resonance position at 528~GHz was reached, which resulted in an optical NEP of 16~pW/$\sqrt{\text{Hz}}$. With this value, the detector performed much better (at its resonance frequency) than substrate-lens-coupled \textit{broadband} TeraFET detectors \cite{Bauer2019, Javadi2021, Ikamas2018}. \textbf{This is the first time that superior performance of a narrow-band TeraFET over broad-band ones can be demonstrated.} This is due to the performance enhancement by the superstrate lens which allows to reap the sensitivity potential inherent in the narrow-band devices. 



We note here, that the simulated directivity of the Case-C sample device was 14~dBi and the radiation efficiency improved from about 20\% to nearly 70\%. Despite these improvements, not all the power available in the beam was fully delivered to the detector. There is hence still room for further improvement. 
\\

We summarize operational data and the achieved performance values of the three devices in Table~\ref{TableFGNR}. The lower sensitivity of the detector in Case B in comparison with Case C is attributed mainly to higher radiation attenuation in the cover layer, paraffin wax exhibiting a non-negligible absorption at THz frequencies.

\begin{table}[!h]
\centering
\caption{Resonance frequency, applied gate bias, and the optical NEP and responsivity values obtained at the respective resonance frequencies.}
\label{TableFGNR}
\begin{tabular}{lcccc}
 &  Frequency& Gate bias  & NEP & $R_{V}$  \\
 &  GHz& V & W$/\sqrt(\text{Hz})$ & V/W  \\
 \hline
 \hline
Case A & 575 &  0.5&  2.5E-11&  528 \\
Case B&  560&  0.5&  2.38E-11& 559  \\
Case C&  528& 0.54&  1.63E-11& 656 
\end{tabular}
\end{table}

\section{Exploitation of the Frequency Shift For a Spectroscopic Application}

As discussed above, the substrate lens assembly brings about a substantial down-shift of the resonance frequency. In principle, the amount of down-shift can be engineered over a considerable frequency range. 

This opens the possibility to optimize or fine-tune a detector for a specific target frequency. We demonstrate this here by using detector pixel ``A1'', which was designed to approximately match the absorption line of ammonia at 600~GHz \cite{Harde01}, for the spectroscopy of lactose with its absorption line at 530~GHz \cite{Brown_2007, DaSilva_2022, Roggenbuck_2010}. For this purpose, we utilized the device of type Case C with its peak sensitivity at 528~GHz.

\begin{figure}[!h]
  \centering  
 \includegraphics[width=0.7\linewidth]{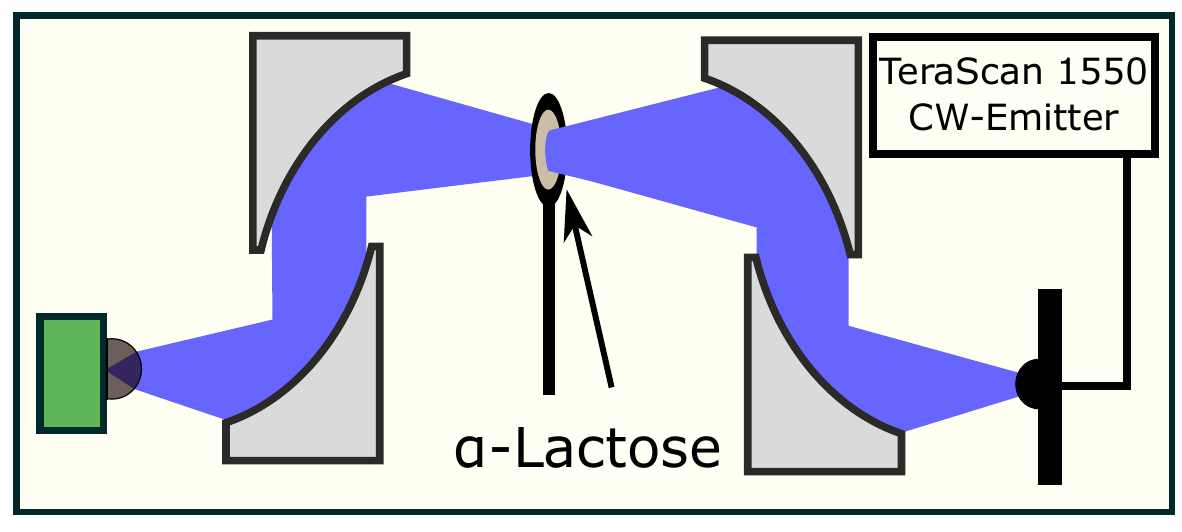}
  \includegraphics[width=\linewidth]{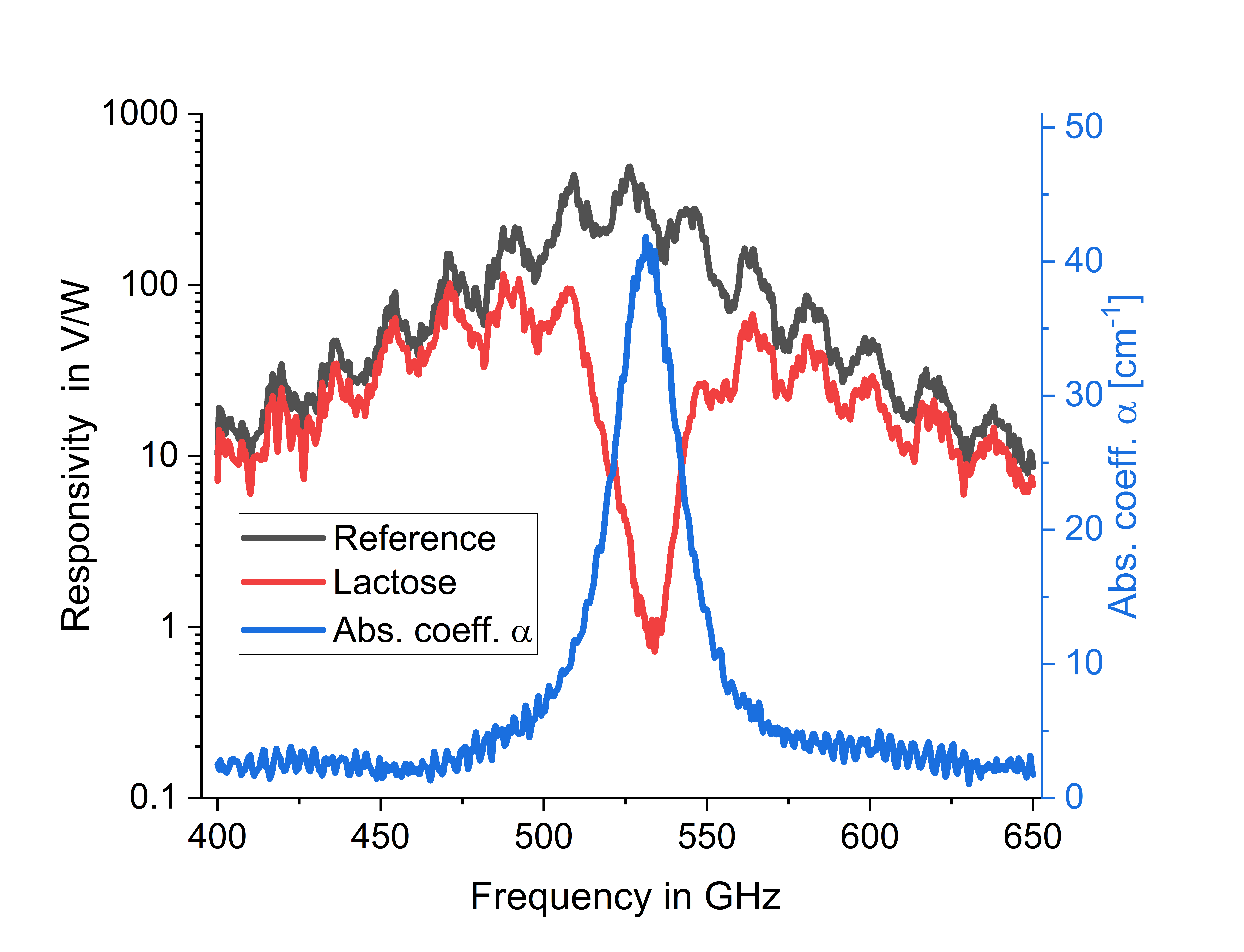}

  \caption{Top: Voltage responsivity of detector with and without $\alpha$-lactose sample and frequency dependent transmission determined from the datasets. Experimentally, a minimum transmission of T(531~GHz)= 0.28\% could be detected above the experimental noise level. Inset-Top: Calculated absorption coefficient $\alpha$ and absorbance. Bottom: Experimental 4-OPA setup for spectroscopy characterization of lactose pellet.}
\label{fig:lactose_measurements_overview}
\end{figure}

For the demonstration experiment, a pellet with commercially available $\alpha$-lactose was prepared. The lactose powder, without a filler material, was mechanically pressed into a 1.4-mm-thick pellet (diameter: 25~mm). It was then placed at the position of the intermediate focus of the measurement system schematically shown in the top panel of Fig.~\ref{fig:lactose_measurements_overview}, and illuminated with the radiation from the emitter of the TeraScan 1550 system (beam power: approx. 1~$\mu$W at 530~GHz).
The bottom panel of Fig.~\ref{fig:lactose_measurements_overview} displays the transmission spectrum (red line) obtained with the sample. For comparison, the grey line shows a null measurement (no sample in the beam path). This curve confirms that the peak responsivity of the detector is indeed close to the absorption line of $\alpha$-lactose.  The oscillations in both curves are due to standing waves in the measurement set-up. With both spectra, we derived the absorption spectrum of $\alpha$-lactose shown as blue curve. 

Measuring with a detector, which is resonant and hence highly sensitive close to the targeted absorption line, ensure a high dynamic range of the measurement. In our experiment, the minimal relative transmittance was determined to be only 2.8\text{\textperthousand} at 531~GHz. The noise floor was not reached yet. 


\section{Appendix}
\begin{figure}[!h]
  \centering
  \includegraphics[width=0.95\linewidth]{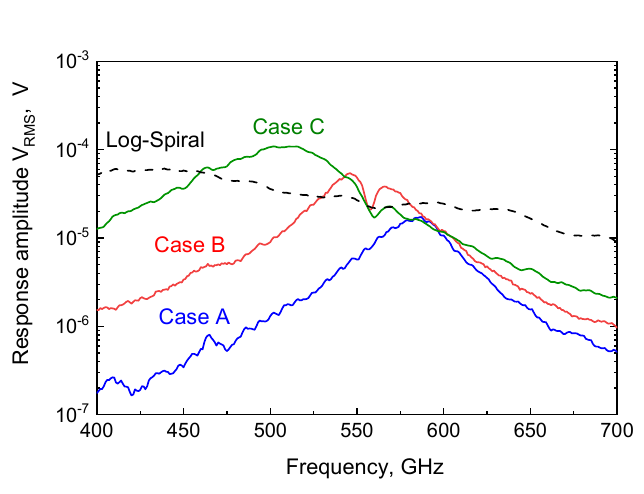}
  \caption {Measured rectified voltage for Case A to C as a function of the radiation frequency (full lines). 
For comparison, the rectified voltage obtained with broad-band detector with a log-spiral antenna is displayed (dashed curve). 
}
\label{fig:image1}
\end{figure}

In Fig.~\ref{fig:image1}, we show additional measurement results of the response of the detectors of the types Case A to Case C to the THz radiation derived from the TeraScan 1550's photomixer emitters. In contrast to the results of Fig.~\ref{RespNep}, unprocessed lock-in read-out voltages (not averaged) are displayed. In addition, the dashed curve displays the response recorded with a broad-band Si CMOS TeraFET which has a fairly flat frequency response in the spectral range covered here. The signal measured with this detector hence represents to a good approximation the power spectrum of the THz emitter.
Going from the data for Case~A to those of Case~B and Case~C, the peak response is measured at ever lower frequency. In that sequence, the measured maximal response amplitude increases faster than the beam's power. This is additional proof for the higher sensitivity in Case C as compared to the other two cases.  Note, that voltage responsivities (defined as the measured rectified voltage divided by the power of the beam), as plotted in Fig.~\ref{RespNep}, are independent of the power of the radiation in the regime of linear response of the detector.


\section*{Acknowledgment}

The German side acknowledges Deutsche Forschungsgemeinschaft (DFG) for the grant no. RO 770/43-1 of the trilateral Fraunhofer-DFG transfer project program as well as RO 770 49-1, a project of the INTEREST collaborative research program. The Lithuanian team is thankful to the Lithuanian Research Council for funding under the contract S-MIP-22-83. D.B. is thankful for funding by European Union (ERC ”TERAPLASM”, project number: 101053716) and the European Union’s Horizon 2020 research and innovation program (grant No 857470). The authors would like to thank Erik Waller from Fraunhofer ITWM Kaiserslautern for SEM pictures of the chip surface.

\bibliographystyle{IEEEtran}

\begin{IEEEbiography}[{\includegraphics[width=1in,height=1.25in,clip,keepaspectratio]{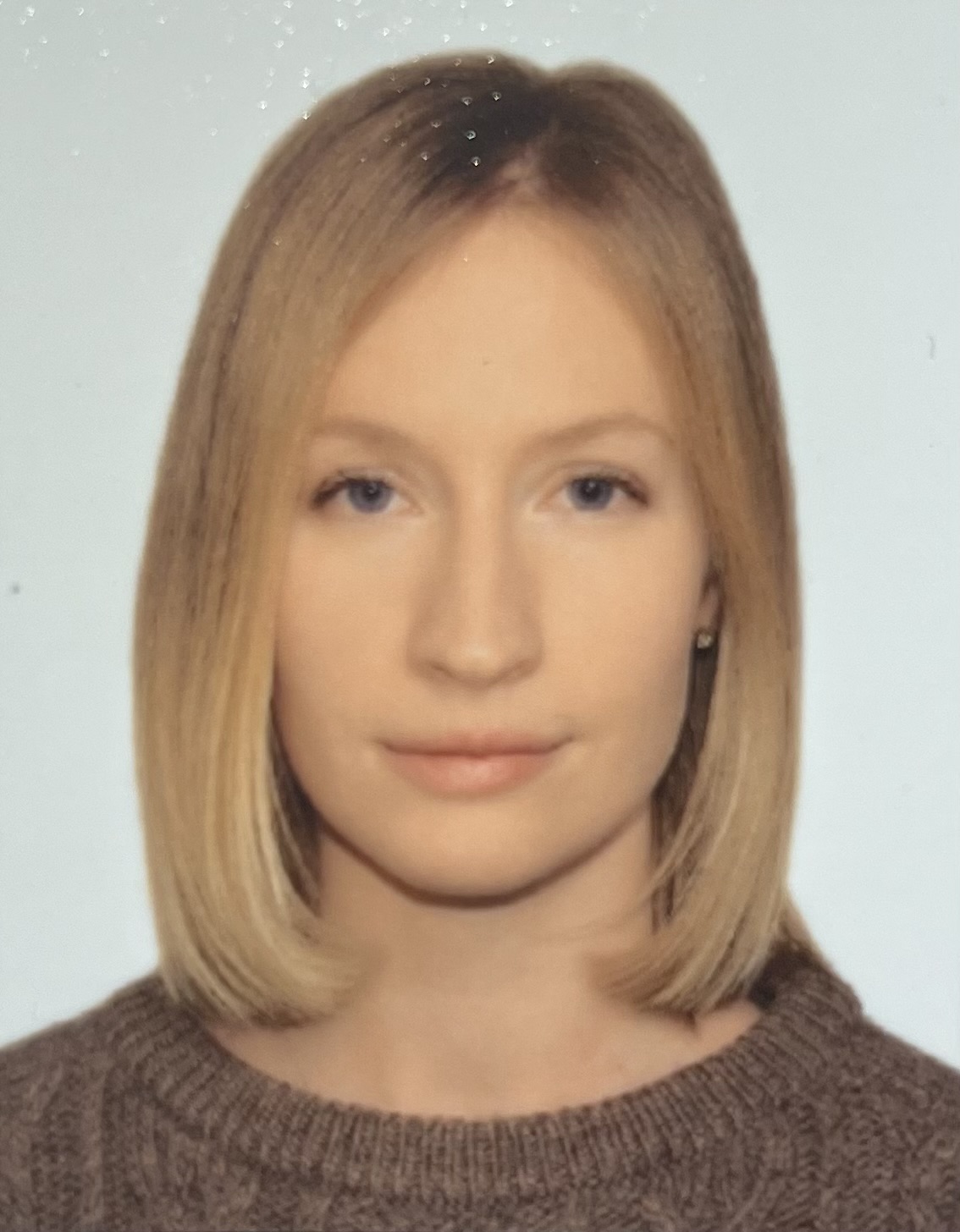}}]
{Anastasiya Krysl} received the Master degree in physical and mathematical sciences at the Belarusian State University in Minsk, Belarus in 2017. She joined the research group of Professor Roskos at the Johann Wolfgang Goethe-Universität Frankfurt am Main as a Ph.D. student in 2019. Her research work is concentrated around design, modeling, characterization and performance optimization of Si CMOS TeraFETs.
\end{IEEEbiography}
\begin{IEEEbiography}[{\includegraphics[width=1in,height=1.25in,clip,keepaspectratio]{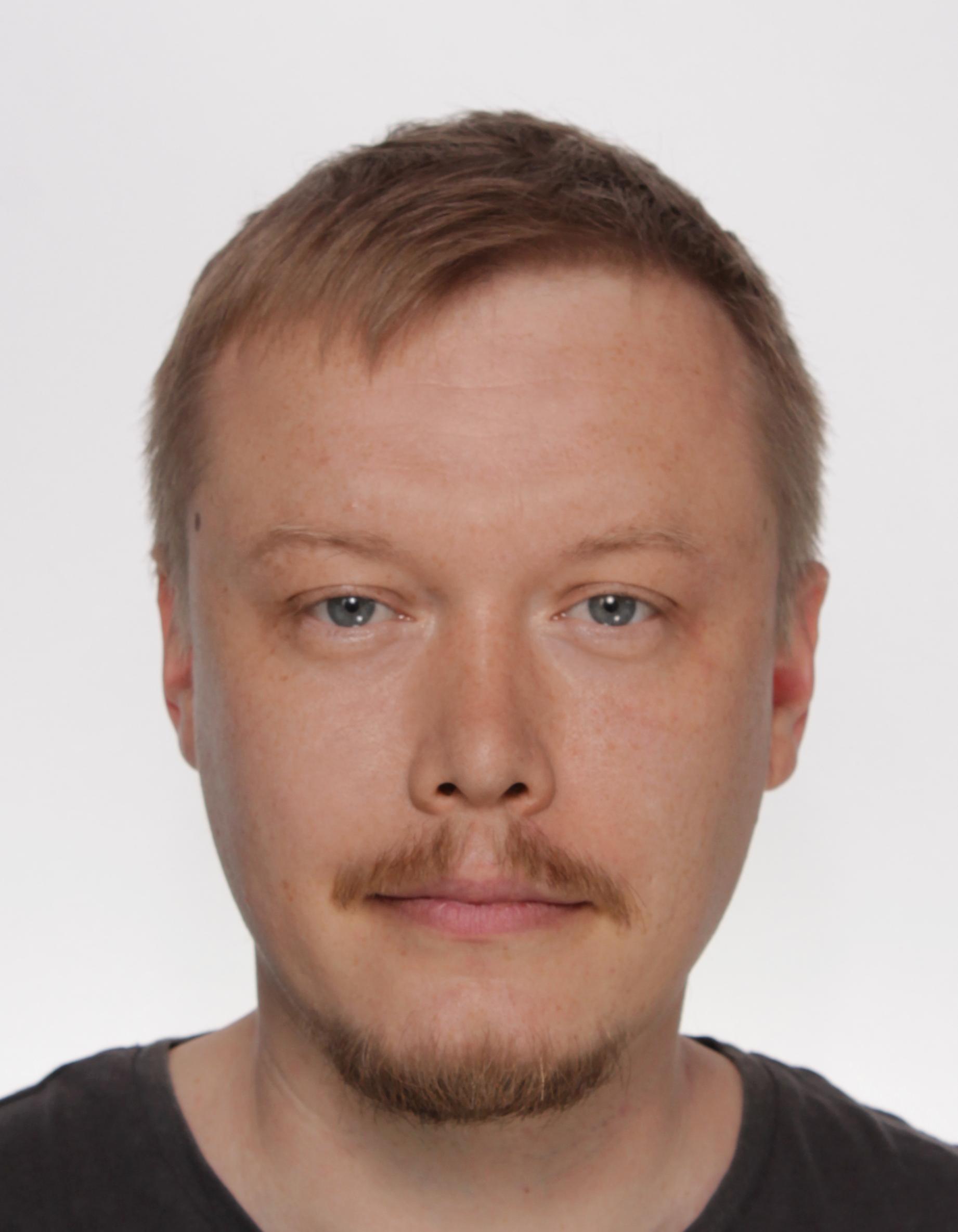}}]
{Dmytro B. But} received the Diploma in microelectronics engineering from the Kyiv Polytechnic Institute NTUU, in Ukraine, in 2009 and received his PhD degree in Physics from the Laboratoire Charles Coulomb, the University of Montpellier II in 2014. In 2014, he also worked as an engineer in the Department of Physics and Technology of Low-dimensional Systems at the V.E. Lashkaryov Institute of Semiconductor Physics NAS of Ukraine. Since 2015 he has continued work as an associate researcher in the group of terahertz spectroscopy at the University of Montpellier. Since 2017, he has been working at the Institute of Electrodynamics, Microwave and Circuit Engineering at the Technical University of Wien as a university assistant. Since 2018, he joined a research group at the Institute of High Pressure Physics of the Polish Academy of Sciences. From 2019, he joined CENTERA Project at the same institution. His research interests include condition matter physics, terahertz electronics, systems and semiconductor devices for them.
\end{IEEEbiography}
\begin{IEEEbiography}[{\includegraphics[width=1in,height=1.25in,clip,keepaspectratio]{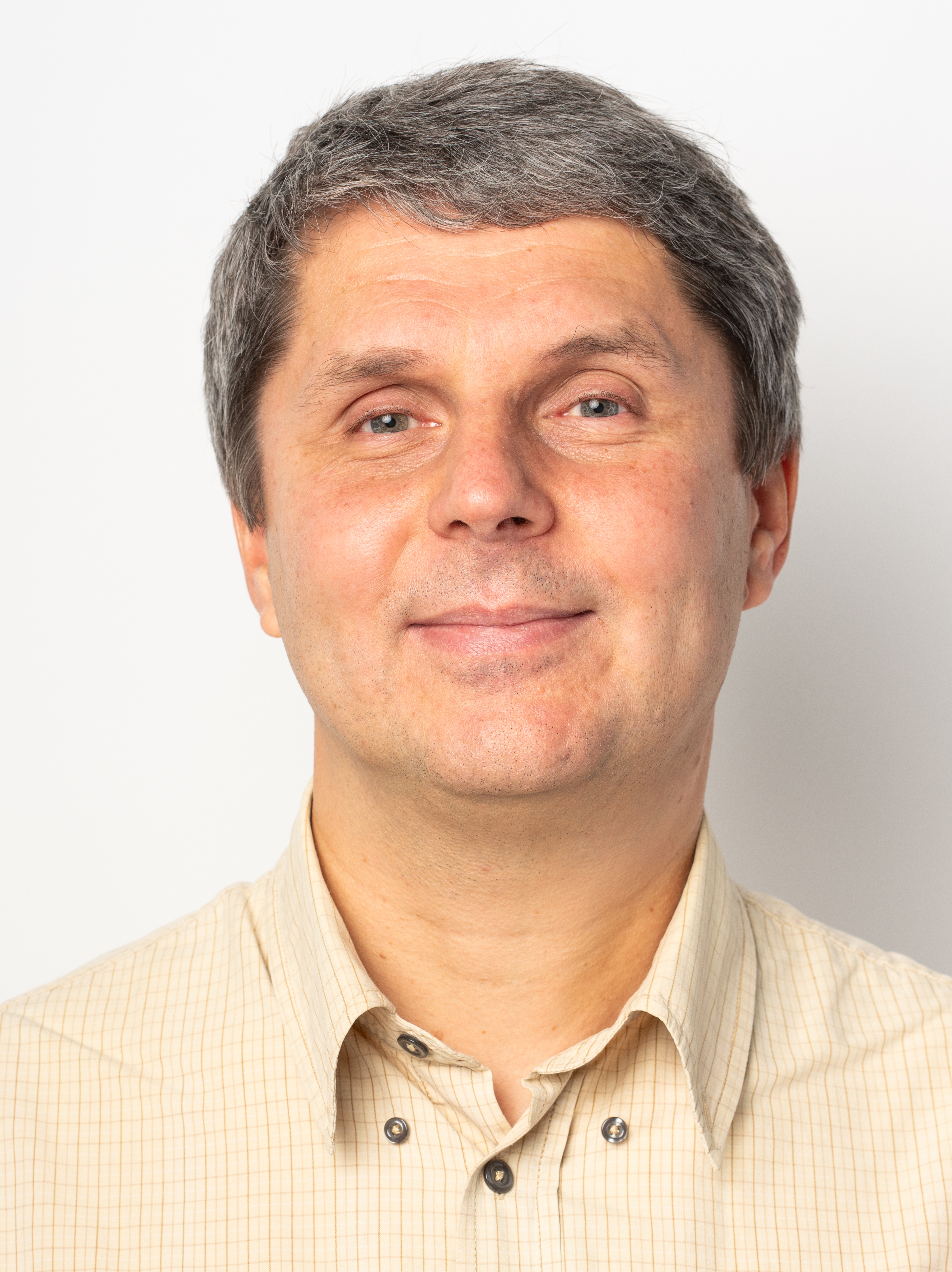}}]
{Kęstutis Ikamas} received the Diploma in physics in 1995 and the doctorate degree working on the modeling of broadband THz detectors with field-effect transistors and the application of these devices for systems with pulsed and dc sources, in 2018, from Vilnius University, Vilnius, Lithuania. He is currently with the Noise and Terahertz Electronics Group, Vilnius University, where he is involved in the area of CMOS transistor-based terahertz detectors’ and sources’ design, modeling, and application.
\end{IEEEbiography}
\begin{IEEEbiographynophoto}{Jakob Holstein}received first state examination diploma (M.Ed.) for mathematics and physics upper secondary class teaching profession in 2020 from Johann Wolfgang Goethe-University Frankfurt am Main, Germany. 
In 2021 and 2022, he received  B.Sc. and M.Sc. degree in physics from the same university. He is a Ph.D. student in the Ultrafast Spectroscopy and
Terahertz Physics Group, Johann
Wolfgang Goethe-Universität Frankfurt am Main, Germany. His research interests focus on experimental characterization of Si-CMOS and graphene based TeraFETs. 
He is also working on integrated TeraFET systems for sensing applications, especially on gas spectroscopy where improvement of THz-power coupling mechanisms plays an important role. 

\end{IEEEbiographynophoto}

\begin{IEEEbiography}[{\includegraphics[width=1in,height=1.25in,clip,keepaspectratio]{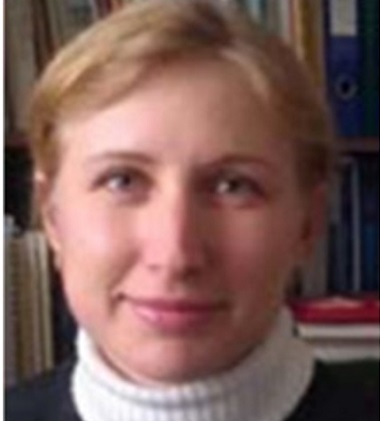}}]
{Anna Shevchik-Shekera} received her Ph.D. degree in technology, equipment and production of electronic equipment in 2019 at National Academy of Sciences of Ukraine (V.E. Lashkaryov Institute of Semiconductor Physics). She is a researcher at V.E. Lashkaryov Institute of Semiconductor Physics. She has authored over 35 publications, 4 patents. The area of her scientific interests includes: infrared and millimeter-wave imaging of active vision systems, design optical components using 3D printing technology and CNC machining.
\end{IEEEbiography}

\begin{IEEEbiography}[{\includegraphics[width=1in,height=1.25in,clip,keepaspectratio]{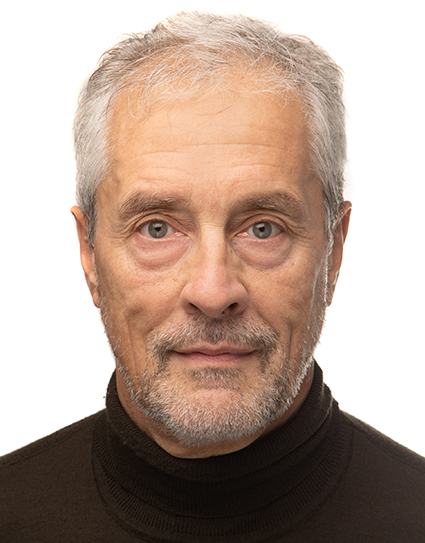}}]{Hartmut G. Roskos}
is Professor of Physics at Johann Wolfgang Goethe-University, Frankfurt am Main, Germany. He studied Physics at the Technical Universities of Karlsruhe and Munich, receiving the PhD degree from the latter in 1989. He then worked at AT\&T Bell Laboratories, Holmdel, USA, where THz phenomena became the focus of his research. He joined the Institute of Semiconductor Electronics of RWTH Aachen in 1991. After receiving the Habilitation degree with a thesis on \textit{Coherent Phenomena in Solid-State Physics Investigated by THz Spectroscopy}, he became a Full Professor at Johann Wolfgang Goethe-Universität in 1997. Current fields of work of his group are time-resolved spectroscopy of solid-state materials (mainly materials with strong electronic correlations), strong light-matter coupling with THz metamaterials, metamaterial sensor development, the development of TeraFETs in various device technologies (including graphene FETs), the development of coherently radiating arrays of resonant-tunneling diodes, THz holographic imaging, and the use of Deep Learning for THz imaging.
He spent sabbatical semesters at the University of California at Santa Barbara in 2005 and at the University of Rochester in 2014, and was an Invited Guest Professor at Osaka University’s Institute of Laser Engineering during the winter semester 2009/2010. In 2009, OC Oerlikon AG awarded his group jointly with the Ferdinand-Braun-Institute (FBH) in Berlin a 5-year endowed professorship which led to the establishment of a Joint Lab for THz Photonics.  
\end{IEEEbiography}
\begin{IEEEbiography}[{\includegraphics[width=1in,height=1.25in,clip,keepaspectratio]{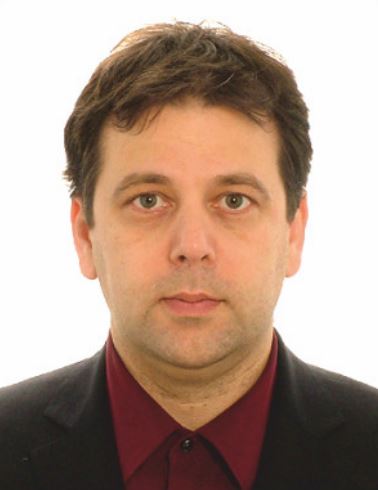}}]{Alvydas Lisauskas}
Prof. IHPP PAS Alvydas Lisauskas received the Diploma in physics from Vilnius University in 1995, and the PhD degree from the Royal Institute of Technology, Stockholm in 2001. In 2002, he joined the Ultrafast Spectroscopy and Terahertz Physics Group at the Goethe University Frankfurt, Germany, working on novel semiconductor devices for THz applications.
From 2014, he is a professor and leading researcher at Vilnius University. From 2014 till 2016 he was a head of joint research laboratory on electrical fluctuations established between the Center for Physical Sciences and Technology and Vilnius University. From 2019 till 2023, he joined CENTERA Project of the Institute of High Pressure Physics PAS, where he led a workgroup on THz electronics.
\end{IEEEbiography}

\end{document}